\documentclass[aps,pra,twocolumn,showpacs,groupedaddress]{revtex4-1}
\usepackage{pifont}
\usepackage{amsmath}
\usepackage{graphicx}
\usepackage{amssymb}
\usepackage{amsfonts}
\usepackage{subfigure}
\usepackage{booktabs}
\usepackage{setspace}
\usepackage{threeparttable}
\usepackage{enumerate}

\begin{document}

\title{Normal state properties of spin-orbit coupled Fermi gases in the
upper branch of energy spectrum}
\author{Xiao-Lu Yu}
\author{Shang-Shun Zhang}
\author{Wu-Ming Liu}
\affiliation{Beijing National Laboratory for Condensed Matter Physics, Institute of
Physics, Chinese Academy of Sciences, Beijing 100190, China }
\date{\today }

\begin{abstract}
We investigate normal state properties of spin-orbit coupled Fermi gases with repulsive $s$-wave interaction, in the absence of molecule formation, i.e., in the so-called ``upper branch''. Within the framework of random phase approximation, we derive analytical expressions for the quasi-particle lifetime $\tau_s$, the effective mass $m_s^*$, and the Green's function renormalization factor $Z_s$ in the presence of Rashba spin-orbit coupling. In contrast to spin-orbit coupled electron gas with Coulomb interaction,  we show that the modifications are dependent on the Rashba band index $s$, and occur in the first order of the spin-orbit coupling strength. We also calculate experimental observable such as spectral weight, density of state and specific heat, which exhibit significant differences from their counterparts without spin-orbit coupling. We expect our microscopic calculations of these Fermi liquid parameters would have the immediate applications to the spin-orbit coupled Fermi gases in the upper branch of the energy spectrum.
\end{abstract}

\maketitle

\section{Introduction}
Motivated by the recent success on the evidence of Stoner ferromagnetism for
the repulsive Fermi gas in the upper branch of the energy spectrum \cite{GBJo},
there has been increasing interest in the nature of uncondensed Fermi
gas (free of molecules) within the repulsively interacting regime \cite{TLHo}, which
naturedly becomes the well-controlled platform for simulating Landau Fermi
liquid. Much of the interest in ultracold atomic gases comes from their amazing
tunability. A wide range of atomic physics and quantum optics technology
provides unprecedented manipulation of a variety of intriguing quantum
phenomena. Based on the Berry phase effect \cite{Berry1984} and its
non-Abelian generalization \cite{Wilczek1984}, Spielman's group in NIST has
successfully generated a synthetic external Abelian or non-Abelian gauge
potential coupled to neutral atoms. Recent experiments have realized the atomic
$^{40}$K \cite{JZhang} or $^6$Li \cite{Zwierlein} gases with spin-orbit coupling (SOC).
These achievements will open a whole new avenue in cold atom physics \cite{NIST1,NIST2,NIST3,NIST4}.

The effect of SOC in fermionic systems has consequently become an important issue in recent years, and attracts a great deal of attentions in ultracold Fermi gases. Most of existing works are devoted to the effect of SOC on the superfluid state with negative $s$-wave scattering length \cite{HZhai2,HHui,RYLiao} and the Bardeen-Cooper-Schrieffer (BCS) to Bose-Einstein condensation (BEC) crossover \cite{Melo,GChen,Vyasanakere,He}. Furthermore, the SOC give rise to a variety of topological phases, such as the quantum spin Hall state and the topological superfluid \cite{Kubasiak,Bermudez,GLiu,JDSau}. In addition to the study of SOC effect on these symmetry-breaking or topological phases, the normal state contains various potential instabilities and deserves attention. The consideration of SOC systems in the framework of Fermi liquid theory is therefore desirable. As is well known, Landau's Fermi-liquid theory provides a phenomenological approach to describe the properties of strongly interacting fermions. Fermi liquid parameters characterizing the renormalized many-body effective interactions have to be determined through experimental results.

The purpose of this paper is to study  key normal-state properties of  two dimensional (2D) Fermi gases with SOC in the repulsive regime---their quasi-particle lifetime, effective mass and Green's function renormalization factor.  Following previous studies of Landau's Fermi-liquid theory including SOC \cite{Fujita,CJWu}, we are attempted to build a microscopic foundation of phenomenological parameters. Therefore within the framework of random phase approximation (RPA), we derive analytical expressions for the quasi-particle lifetime $\tau_s$, the effective mass $m^*_s$, and the Green's function renormalization factor $Z_s$ for a 2D Fermi gases with repulsive $s$-wave interaction in the presence of Rashba SOC. To make contact with current experiments directly, we also calculate experimental observable such as spectral weight, density of state and specific heat, and discuss their corresponding experimental signatures. We shall show that the modifications are dependent on the Rashba band index $s$ denoting the two directions of the eigenspinors of the Rashba Hamiltonian.

The paper is organized as follows. The model Hamiltonian and the renormalizations due to $s$-wave interaction in the presence of SOC is discussed in Sec. II. In Sec. III, Starting from the RPA self-energy of the SOC Fermi liquid, we derived all the analytical formula of the Fermi liquid parameters in presence of SOC. The experimental observable quantities such as the spectral function, density of states and specific heat are calculated. Sec. IV is devoted to discuss the experimental measurements of these fundamental parameters and their corresponding experimental signatures. The comparisons with the ordinary Fermi liquid are also presented.

\section{pertubative theory of 2D Fermi gases with $s$-wave interaction in the
presence of Rashba SOC}

\subsection{Model Hamiltonian}

We consider a 2D spin-1/2 fermionic system with Rashba-type SOC and $s$-wave interaction, described by the model Hamiltonian
\begin{equation}
\mathcal{H}=\mathcal{H}_0+\mathcal{H}_{I}.
\end{equation}
The non-interacting part $\mathcal{H}_0$ reads as,
\begin{equation}
\mathcal{H}_0=\sum\limits_{\mathbf{p}} c^{\dag}_{\mathbf{p}}[\frac{\mathbf{p}^{2}}{2m}+\alpha\left( \mathbf{\hat{z}}\times \mathbf{p}\right) \cdot \mbox{\boldmath$\sigma$} -\mu]c_{\mathbf{p}},
\end{equation}
where $c_{\mathbf{p}}=(c_{\mathbf{p},\uparrow},c_{\mathbf{p},\downarrow})^T$, $\mu=k_{F}^{2}/2m$ is the chemical potential, $\alpha $ represents SOC strength and $k_F$ is the Fermi momentum in the absence of Rashba-type SOC. The reduced Plank constant $\hbar$ is taken as $1$ in this paper. The non-interacting Hamiltonian $\mathcal{H}_0$ can be diagonalized in the helicity bases
\begin{equation}  \label{eigenstates}
|\mathbf{k},s\rangle =\frac{1}{\sqrt{2}}\left(
\begin{array}{c}
1 \\
ise^{i\phi (\mathbf{k}) }%
\end{array}%
\right) ,s=\pm 1,
\end{equation}
where $\phi(\mathbf{k})=\text{arctan}(k_y/k_x)$ and $s$ is the helicity of Fermi surfaces, which represents that the in-plane spin is right-handed or left-handed to the momentum. The dispersion relations for two helical branches are
$\xi _{\mathbf{k},s}=(\mathbf{k}^{2}+2sk_{R}|\mathbf{k}|-k_{F}^{2})/2m$,
where $k_{R}=m\alpha$ corresponds to the recoil momentum in experiments \cite{JZhang,Zwierlein}. The Fermi surfaces are given by $\xi _{\mathbf{k},s}=0$, which yields two Fermi momenta $k_{s}=\kappa k_F -sk_{R}$ with $\kappa =\sqrt{1+\gamma ^{2}}$. We have defined the dimensionless SOC strength $\gamma=k_{R}/k_{F}$.  Recently, the experimental realization of the SOC degenerated Fermi gases have been reported \cite{NIST4,JZhang,Zwierlein}. By applying a pair of laser beams to the ultracold $^{40}$K or $^{6}$Li atoms trapped in a anisotropic harmony trap, the equal weight combination of the Rashba-type and Dresselhaus-type SOC is generated. Their elegant experiments are performed in the weakly repulsive regime, which could provide the possibilities to study the SOC degenerated Fermi gases in the normal state.

The interacting part reads as
\begin{equation}
\mathcal{H}_{I}=2g\int \frac{d^2\mathbf{k}d^2 \mathbf{p}d^2 \mathbf{q}}{(2\pi)^6}c_{\mathbf{k}+\mathbf{q},\uparrow}^{\dag} c_{\mathbf{p}-\mathbf{q},\downarrow }^{\dag}
c_{\mathbf{p},\downarrow } c_{\mathbf{k},\uparrow },
\end{equation}
where $g=2\pi N a_s/3\sqrt{2 \pi} m \zeta_z$, which is controlled by the $s$-wave scattering length $a_s$. Here $N$ is the total atom number, $\zeta_z=\sqrt{1/m\omega_z}$ is the confinement scale of the atomic cloud in the $\mathbf{\hat{z}}$ direction perpendicular to the 2D plane, and $m$ is the mass for the ultracold atoms. Notice that the universal properties of the low-energy interaction among ultra-cold atoms depend only on the scattering length $a_s$ \cite{Mott,Messiah,Bratten,Kohler,Chin,Inouye,Courteille,Roberts,Vuletic}. We focus on the normal state regime of Fermi atomic gas in this work assuming positive scattering length, which could be reached in the upper branch of Feshbach resonance. We note that the gas of dimers and the repulsive gas of atoms represent two different branches of the many-body system, both corresponding to positive values of the scattering length \cite{Castin}. The atomic repulsive gas configuration has been experimentally achieved by ramping up adiabatically the value of the scattering length, starting from the value $a=0$ \cite{Salomon}.

\subsection{Renormalizations due to the $s$-wave interaction.}

Let's consider the problem in the helicity bases $|\mathbf{k},s\rangle$. The non-interacting Green's function is
\begin{equation} \label{eq:G0}
G_s^0(\mathbf{k},\omega)=\frac{1}{\omega-\xi_{\mathbf{k},s}+i\text{sgn}(\omega)0^+}.
\end{equation}
The Dyson's equation expresses the relation between the non-interacting and interacting Green's functions in terms of the self-energy $\Sigma_s$ as
\begin{equation}
G_{s }(\mathbf{k},\omega) =\frac{1}{\omega -\xi _{\mathbf{k},s }-\Sigma _{s}( \mathbf{k},\omega) }.
\end{equation}
All the many-body physics is contained in the self-energy $\Sigma_s$. The poles of interacting Green's function $G_{s }(\mathbf{k},\omega)$ give the quasi-particle excitations,
of which lifetime $\tau_s=1/\Gamma_s$ can be obtained from the imaginary part of self-energy as
\begin{equation}\label{eq:Gammas}
\Gamma _{s }(\mathbf{k}) =-2\text{Im} \Sigma _{s} ( \mathbf{k},\xi_{\mathbf{k},s}).
\end{equation}

The real part of the self-energy gives a modification of dispersion relations. At low temperature, the properties of the low energy excitation in the vicinity of the Fermi surface is essential. Thus we can expand the real part of self-energy to the first order of $\omega $ and $|\mathbf{k}|-k_s$ as
\begin{align}
\text{Re}\Sigma _{s }( \mathbf{k},\omega) &=\text{Re}\Sigma _{s }( \mathbf{k}_s,0) +\omega
\partial _{\omega }\text{Re}\Sigma _{s}( \mathbf{k}_s,\omega)|_0 \notag \\
&+(\mathbf{k}-\mathbf{k}_s) \cdot \nabla_{\mathbf{k}}\text{Re}\Sigma _{s }(\mathbf{k},0)|_{\mathbf{k}_s}.
\end{align}
The interacting Green's function now becomes
\begin{equation}
G_{s }\left( \mathbf{k},\omega \right) = \frac{Z_s}{\omega -\xi_{\mathbf{k},s}^*
+i(1/2)\Gamma_s(\mathbf{k}) },
\end{equation}%
where $\xi_{\mathbf{k},s}^*$ is the modified energy dispersion. The Green's function acquires a renormalized factor
\begin{equation}\label{eq:Zs}
Z_{s }=\frac{1}{1-A_s},
\end{equation}
where $A_s=\partial _{\omega }\text{Re}\Sigma _{s }( \mathbf{k}_s,\omega)|_0$. The Fermi velocity $v_{k_s}=\partial \xi^*_{\mathbf{k},s}/\partial k|_{k_s}$ can be calculated as follows
\begin{equation}  \label{eq:xirenorm}
\frac{\partial\xi _{\mathbf{k},s}^{\ast }}{\partial k}|_{|\mathbf{k}|=k_s} = Z_s(\kappa k_F/m+\partial _{k}\text{Re}\Sigma _{s
}( \mathbf{k},0)|_{k_s}).
\end{equation}
For non-interacting Fermi gas, the Fermi velocity is $v_{k_s}^0=\kappa k_F/m$. Therefore we introduce the effective mass via $\partial \xi^*_{\mathbf{k},s}/\partial k |_{|\mathbf{k}|=k_s}=\kappa k_F/m^*_s$. The effective mass in terms of self-energy is
\begin{equation}  \label{eq:meff}
\frac{m^{\ast }_{s}}{m}=\frac{1}{Z_{s }}\left( 1+\frac{m}{\kappa k_F}\partial
_{k}\text{Re}\Sigma _{s }^{R}\left( \mathbf{k}_{s },0\right) \right) ^{-1}.
\end{equation}

The Eqs. (\ref{eq:Gammas}), (\ref{eq:Zs}) and (\ref{eq:meff}) are our starting points of microscopic calculations of normal states properties, which embody the main properties of a quasi-particle in the Landau theory of Fermi liquids.

\begin{figure}[!t]
\begin{center}
\includegraphics[width=2.8in]{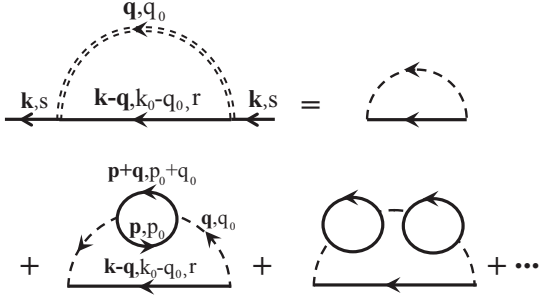}\hspace{0.5cm}
\end{center}
\caption{ Feynman diagrams for the self-energy of the SOC Fermi liquid in the presence of $s$-wave interaction. The Feynman rules are defined under the helicity bases. The label $s$ and $r$ denote the helicity index. The self-energy is calculated within the framework of RPA \cite{Negele}.}
\label{fig:self-energy}
\end{figure}

\section{SOC Fermi liquid parameters with repulsive $s$-wave interaction}

\subsection{RPA Self-energy}

To investigate the renormalization effects in two Rashba energy bands separately, it is convenient to work in the helicity bases. The interacting part of the Hamiltonian in the helicity bases is
rewritten as
\begin{eqnarray}
\hspace{-0.1cm}\mathcal{H}_{I} = \hspace{-0.2cm} \sum\limits_{\mathbf{k,p,q}} V_{s s^\prime;r r^\prime}(\mathbf{k,p,q}%
)\varphi _{\mathbf{k+q},s^\prime}^{\dag } \varphi _{\mathbf{p-q}%
,r^\prime}^{\dag } \varphi _{\mathbf{p},r} \varphi _{ \mathbf{k},s},
\end{eqnarray}%
where the interaction vertex $V_{s s^\prime;r r^\prime}(\mathbf{k,p,q}%
)=gf_{ss^{^{\prime }}}(\theta _{\mathbf{k}},\theta _{\mathbf{k+q}%
})f_{rr^{^{\prime }}}(\theta _{\mathbf{p}},\theta _{\mathbf{p-q}})$. Due to
the presence of SOC, the spin is locked to momentum. The interaction vertex
acquires an overlap factor $f_{ss^{^{\prime }}}(\theta _{\mathbf{k} },\theta _{\mathbf{%
k+q}})$ and $f_{rr^{^{\prime }}}(\theta _{\mathbf{p}},\theta _{ \mathbf{p-q}%
})$, which is defined by
\begin{eqnarray}\label{eq:overlap}
f_{ss^{^{\prime }}}(\theta _{\mathbf{k}},\theta _{\mathbf{p}})= \frac{1}{2}%
(1+ss^{^{\prime }}e^{i\left[ \theta _{\mathbf{k}} -\theta _{\mathbf{p}} %
\right] }),
\end{eqnarray}%
where $\theta _{\mathbf{k}}$ and $\theta _{\mathbf{p}}$ are the azimythal angles of $\mathbf{k}$ and $\mathbf{p}$ respectively.
 Within the framework
of RPA, the interaction vertex is modified as (see Fig. \ref{fig:self-energy})
\begin{equation}
V_{ss^{^{\prime }},rr^{^{\prime }}}^{RPA}(\mathbf{k,p,q},\omega)=\frac{g}{\epsilon (\mathbf{q},\omega )}
f_{ss^{^{\prime }}}(\theta _{\mathbf{k}},\theta _{\mathbf{k+q}
})f_{rr^{^{\prime }}}(\theta _{\mathbf{p}},\theta _{\mathbf{p-q}}),
\end{equation}
where the dielectric function $\epsilon (\mathbf{q},\omega )=1+g\chi (\mathbf{q},\omega )$ and $\chi (\mathbf{q},\omega )$ is the bare density-density susceptibility of non-interacting SOC Fermi gas. In the long wavelength and low frequency limit, the susceptibility can be carried out as
\begin{eqnarray}\label{eq:chi0}
\text{Re}\chi (y) &=&\frac{m}{\pi }\left[ 1-\frac{|y|}{\sqrt{y^{2}-1}}\Theta
(\left\vert y\right\vert -1)\right] , \notag \\
\text{Im}\chi (y) &=&\frac{m}{\pi }\frac{y}{\sqrt{1-y^{2}}}\Theta (1-\left\vert
y\right\vert ),
\end{eqnarray}
where $y=m\omega/\kappa k_F |\mathbf{q}|$. It can be seen from Eq. (\ref{eq:chi0}) that the susceptibility
satisfies the following relations
\begin{figure}[!t]
\centering
\includegraphics[width=3.1in]{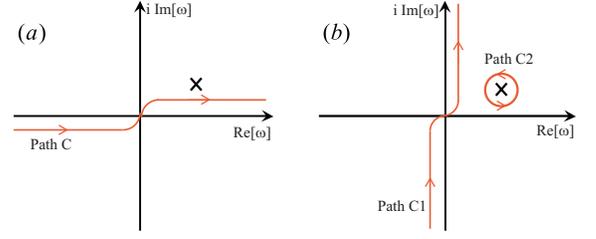}
\caption{(Color online). (a): The contours of integration in the complex $%
\protect\omega$ plane for the self-energy given by Eq. (\ref{self-energy}).
For intermediate states with the energy $0<\protect\xi _{\mathbf{k-q,}r}<\protect\omega$,
the pole of the propagator falls in the first quadrant. (b): Schematic of the
deformation of the contour into the imaginary axis. The self-energy given by Eq. (\ref{self-energy})
is equal to the integration along the path C1 with an additional contribution of the residue
as shown by the integral path C2.}
\label{path}
\end{figure}
\begin{equation}\label{eq:relation1}
\chi^*( \mathbf{q},\omega) =\chi( -\mathbf{q},-\omega)=\chi( \mathbf{q},-\omega).
\end{equation}
 It's important to notice that  the bare susceptibility is real for $\left\vert \omega \right\vert >v_{F}\sqrt{1+\gamma ^{2}}|\mathbf{q}|$. For $\left\vert \omega \right\vert <v_{F}\sqrt{1+\gamma ^{2}}|\mathbf{q}|$, the bare susceptibility in Eq. (\ref{eq:chi0}) contains an imaginary part which represents the absorptive behavior of the medium. This imaginary part is responsible for the finite lifetime of the quasi-particle in the medium. Along the imaginary axis, the analytical formula of $\chi(\mathbf{q},\omega)$ is much simpler (see Fig. \ref{path}) as
\begin{equation}\label{eq:chi2}
\chi (\mathbf{q},i\omega)=\frac{m}{\pi }[ 1-\frac{|y|}{\sqrt{y^{2}+1}}].
\end{equation}
The RPA self-energy (see Fig. \ref{fig:self-energy}) is
\begin{equation}
\Sigma _{s}(\mathbf{k},\omega)=i\int_{C}\frac{d^2\mathbf{q}dq_{0}}{(2\pi )^{3}}%
\sum\limits_{r}\frac{g\mathcal{F}_{sr}}{\epsilon (\mathbf{q},q_0 )}G^0_r(\mathbf{k-q},\omega-q_0),
\label{self-energy}
\end{equation}
where the $\mathcal{F}_{sr}$ is the overlap factor
\begin{eqnarray} \label{eq:factor}
\mathcal{F}_{sr} =\frac{1+sr\cos \left( \theta _{\mathbf{k}}-\theta _{\mathbf{k-q}}\right)}{
2}.
\end{eqnarray}
The integral path $C$ is shown in Fig. \ref{path} (a). After deformation of contour path in Fig. \ref{path} (b),
the pole gives rise to a residue contribution
\begin{equation}  \label{residue}
\Sigma _{s}^{\text{pole}}(\mathbf{k},\omega )=-\sum\limits_{r}\int_{D_{r}}\frac{d^2\mathbf{q}
}{(2\pi )^{2}}\frac{g}{\epsilon }\mathcal{F}_{sr},
\end{equation}
where the region of integration $D_{r}$ is $k_{r}<\vert \mathbf{k-q}\vert <\vert \mathbf{k} \vert -(r-s)k_{R}$
and $\epsilon =1+g\chi _{0}\left( \mathbf{q},\omega -\xi _{\mathbf{k}-\mathbf{q},r}\right)$.
The line integral along the imaginary axis is
\begin{equation}  \label{imagline}
\Sigma _{s}^{\text{line}}(\mathbf{k},\omega )=-\int_{-\infty }^{\infty }\frac{d^2\mathbf{q}dq_0
}{(2\pi )^{3}}\sum\limits_{r}\frac{g\mathcal{F}_{sr}}{\epsilon (\mathbf{q},iq_0)}\frac{1}{\omega -iq_0-\xi _{\mathbf{k}-\mathbf{q}r}}.
\end{equation}
The total self-energy is given by
\begin{equation}
\Sigma _{s}=\Sigma _{s}^{\text{pole}}+\Sigma _{s}^{\text{line}}.
\end{equation}
The advantage of this decomposition is that the imaginary part is given
by the contribution of residue, and the real part is mainly determined by the
line integral.

\subsection{Quasi-particle lifetime}

The quasi-particle lifetime can be calculated by the imaginary part
of self-energy, which comes from the contribution of residue in Eq. (\ref{residue}).
At zero temperature and for quasi-particle $\xi _{\mathbf{k},s}>0$, the
imaginary part of the self-energy reads
\begin{align} \label{ltime}
\Gamma_{s}(\mathbf{k})=
&2{\displaystyle \sum \limits_{\mathbf{q},r}}
\Theta(\xi_{\mathbf{k-q},r})\Theta(\xi_{\mathbf{k,}s}-\xi_{\mathbf{k-q}
,r})\nonumber\\
&\times \text{Im}V^{RPA}_{sr;sr}(\mathbf{k,k-q,q}
;\xi_{\mathbf{k,}s}-\xi_{\mathbf{k-q},r}).
\end{align}

\begin{figure}[!t]
\centering
\includegraphics[width=2.9in]{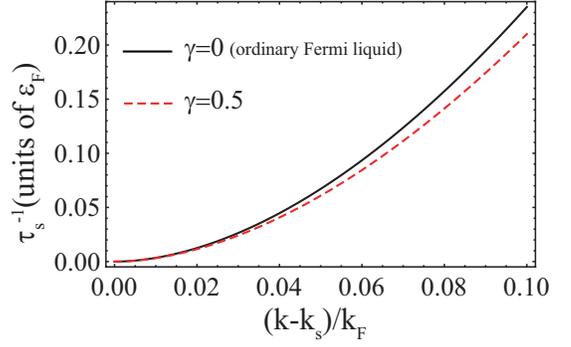}
\caption{(Color online). The inverse of the lifetime $\tau_s$ for $^{40}$K ultracold atoms as a function of the momentum $k$ in the vicinity of the Fermi surface. The lifetime of quasi-particle is enhanced due to the presence of SOC. The parameters taken here are: the number of atoms is about $10^{4}$, $k_R=h/\lambda$ with $\lambda=773$nm, $\gamma=0.5$, trap frequency $\omega_z= 2 \pi \times 400$Hz, and $a_s=32 a_0$, where $a_0$ is the Bohr Radius. The unit $\epsilon_F=\hbar \times 0.21$MHz.}
\label{figure:lifetime}
\end{figure}

The imaginary part of the RPA vertex in the helicity bases is
\begin{equation}
\text{Im}V_{sr;rs}^{RPA}\left( \mathbf{k,k-q,q,k};w\right) =g\mathcal{F}_{sr}\text{Im}\frac{1}{%
\epsilon (y) },
\end{equation}%
where $w=\xi _{\mathbf{k},s}-\xi _{\mathbf{k-q},r}$ and $y=w/v_{F}|\mathbf{q}|$. Since the main contribution of integral comes from the forward scattering, i.e., $y\ll 1$, the density-density susceptibility can be expanded about $y=0$. One finds that the susceptibility in this neighborhood is
\begin{equation}
\chi(y)=\frac{m}{\pi }\left( 1+iy\right) +\mathcal{O}(y^{2}).
\end{equation}
The imaginary part of RPA vertex goes to
\begin{equation}  \label{VIm}
\text{Im}V_{sr;rs}^{RPA}\left( \mathbf{k,k-q,q};w\right) \simeq -\frac{m^2g^{2}}{( m g+\pi)^{2}}\frac{\pi }{m}y\mathcal{F}_{sr}.
\end{equation}
Substituting Eq. (\ref{VIm}) into Eq. (\ref{ltime}), the inverse lifetime $\Gamma _{s}(k)$ can be evaluated as
\begin{equation} \label{eq:lifetime}
\Gamma _{s}(\mathbf{k})=-\frac{m^2g^{2}\epsilon_F}{\pi (mg+\pi )^2} \delta^2 \left\{\ln \frac{\delta}{8}- \frac{1}{2}-\gamma^2 \ln \frac{\gamma}{4}\right\},
\end{equation}
where $\delta=(k-k_s)/k_F$. The result for ordinary Fermi liquid in the presence of $s$-wave repulsive interaction can be obtained by taking the limit $\gamma \rightarrow 0$ in Eq. (\ref{eq:lifetime}). The effect of SOC is shown in Fig. \ref{figure:lifetime} via the comparison with the ordinary Fermi liquid with the same strength of $s$-wave repulsive interaction, where we can conclude that the quasi-particle is much stabler in the presence of SOC.

\subsection{Green's function renormalization factor}

Our starting point is the real part of the self-energy which contains two
parts: one is the residue contribution given by Eq. (\ref{residue}), the
other is the integral along the imaginary axis given by Eq. (\ref{imagline}%
). Thus we want to evaluate%
\begin{eqnarray}\label{eq:As2}
A_{s} &=&A_s^{\text{\text{pole}}}+A_s^{\text{line}} \notag \\
&=&\partial _{\xi }\Sigma _{s}^{\text{pole}}(\mathbf{k}_{s},\xi )|_{\xi =0}+\partial
_{\xi }\Sigma _{s}^{\text{line}}(\mathbf{k}_{s},\xi )|_{\xi =0}.
\end{eqnarray}
Given the rotation symmetry, we only need to consider $\mathbf{k}_s=k_{s}\mathbf{e}%
_{x}$. The first term in Eq. (\ref{eq:As2}) is
\begin{eqnarray}\label{eq:As_pole}
A_{s}^{\text{pole}} = \sum\limits_{r}\int \frac{d^2 \mathbf{q}}{(2\pi )^{2}}\delta (\xi _{\mathbf{k}_s-\mathbf{q},r})\frac{g}{\epsilon }\mathcal{F}_{sr}.
\end{eqnarray}
The second term in Eq. (\ref{eq:As2}) can be integrated by parts, which gives two parts: the first one reads as
\begin{equation}
A_{s}^{\text{line}}=-\sum\limits_{r}\int \frac{d^2 \mathbf{q}}{(2\pi )^{2}}\delta (\xi _{\mathbf{k}_s-\mathbf{q},r})\frac{g}{\epsilon }\mathcal{F}_{sr},
\end{equation}
which cancels the residue contribution given by Eq. (\ref{eq:As_pole}). Thus the final result of Eq. (\ref{eq:As2}) can be expressed as
\begin{equation}
A_{s}=-\frac{mgk_F^2}{(2\pi )^{3}k_{s}^{2}}\int_{0}^{2\pi }d\phi \int_{0}^{\infty} d \bar{y} \int_{0}^{\infty }dx \sum\limits_{r}\text{Im} f_{sr}(x,\bar{y},\phi ),  \label{As}
\end{equation}
where we have defined $\bar{y}=mw/qk_{s}, x=q/2k_{s}$ and
\begin{equation}\label{eq:As}
f_{sr}(x,\bar{y},\phi )=\frac{\mathcal{F}_{sr}}{
i\bar{y}-\mu (x,\phi )}\frac{1}{\epsilon
^{2}}\frac{\partial \epsilon }{\partial \bar{y}}.
\end{equation}
Here $\mu_{s,r} (x,\phi )=m \xi_{\mathbf{k-q},r}/|\mathbf{q}|k_s$, and the overlap factor $\mathcal{F}_{sr}(x,\phi )=1/2+sr(1-2x\cos \phi )/2l$, with $l=|\mathbf{k-q}|/k_{s}=\sqrt{1-4x\cos \phi+4x^{2}}$.
\begin{figure}[t]
\centering
\includegraphics[width=2.8in]{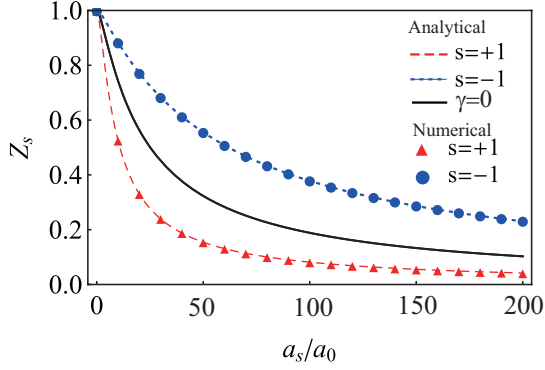}
\caption{(Color online). Renormalization factor $Z_s$ as functions of scattering length $a_s$ with the same parameters in Fig. \ref{figure:lifetime}. The black thick line represents the renormalization factor of ordinary Fermi gases ($\gamma=0$). The red (dashed) and blue (dotted) lines represent the analytical results for the SOC Fermi gas given by Eq. (\ref{resultZs}). The discrete points are evaluated numerically.}
\label{figure:Zs}
\end{figure}

\begin{figure}[t]
\centering
\includegraphics[width=2.8in]{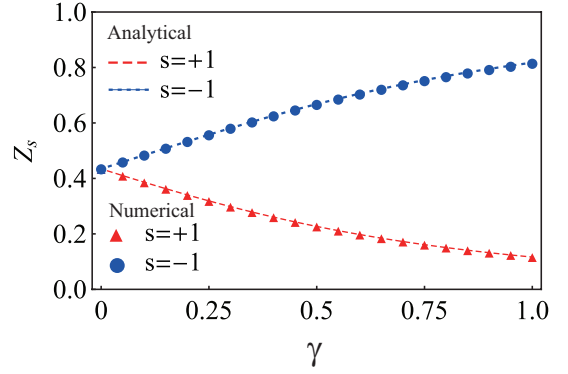}
\caption{(Color online). Renormalization factor $Z_s$ as functions of dimensionless SOC strength $\gamma$ with the same parameters in Fig. \ref{figure:lifetime}. The red (dashed) and blue (dotted) lines represent the analytical results for the SOC Fermi gas given by Eq. (\ref{resultZs}). The discrete points are evaluated numerically.}
\label{figure:Zs2}
\end{figure}

From Eqs. (\ref{eq:Zs}), (\ref{As}) and (\ref{eq:As}), we can obtain the renormalization factor straightforwardly
\begin{eqnarray}\label{resultZs}
Z_s^{-1}&& =  1+\frac{m^2g^{2}}{8\pi \left( mg+\pi \right)}\frac{1}{(\kappa-s \gamma)^{2}} \notag \\
&& = 1+\frac{m^2g^{2}}{8\pi \left( mg+\pi \right)}(1+ s\gamma +\mathcal{O}(\gamma^2)).
\end{eqnarray}
The renormalization factor turns out to be dependent on the helicity $s$, and the leading correction due to SOC is $\mathcal{O}(\gamma)$, which is different from the results of two dimensional electron gas (2DEG) with SOC in semiconductors \cite{Saraga}. We show the analytical results along with the numerical calculation for $Z_s$ as a function of the $s$-wave scattering length in Fig. \ref{figure:Zs} and the strength of SOC in Fig. \ref{figure:Zs2}. The Green's function renormalization factor for ordinary Fermi liquid is also shown for comparison in Fig. \ref{figure:Zs}. We can see from Fig. \ref{figure:Zs2} that the renormalization factor is reduced for the $s=+1$ branch while enhanced for the $s=-1$ branch with increasing strength of SOC.

\subsection{Effective mass}

\begin{figure}[!t]
\centering
\includegraphics[width=3.2in]{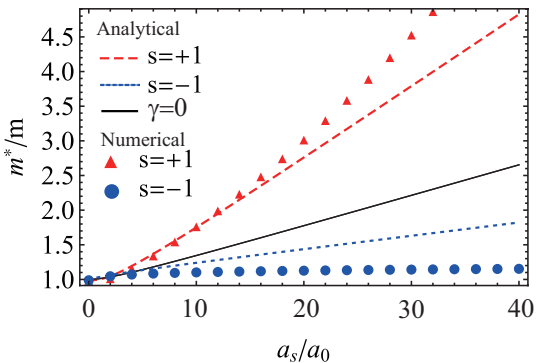}
\caption{(Color online). Effective mass as functions of $s$-wave scattering length $a_s$ with the same parameters as in Fig. \ref{figure:lifetime}. The red (dashed) and blue (dotted) line denote the case for $s=+1$ and $s=-1$ respectively. The solid triangle and circle points are the corresponding numerical results. The black line represents the analytical results for ordinary Fermi liquid ($\gamma=0$). The many-body modifications of the effective mass are dependent on the helical bands.}
\label{figure:meff}
\end{figure}

\begin{figure}[!t]
\centering
\includegraphics[width=2.73in]{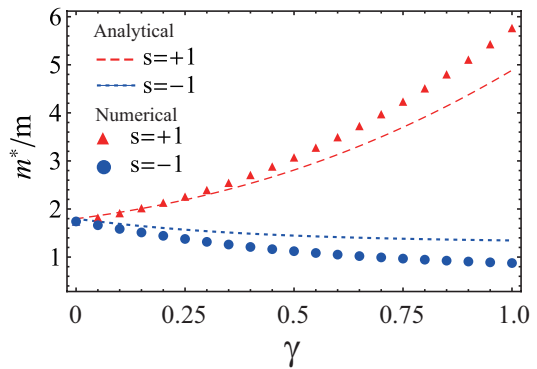}
\caption{(Color online). Effective mass as functions of the strength of SOC $\gamma$ with the same parameters as in Fig. \ref{figure:lifetime}. The red (dashed) and blue (dotted) line denote the case for $s=+1$ and $s=-1$ respectively. The solid triangle and circle points are the corresponding numerical results.}
\label{figure:meff2}
\end{figure}

The effective mass can be evaluated by the real part of the static self-energy in the vicinity of the Fermi surface from Eq.(\ref{eq:meff}). In contrast to the calculation of the renormalization factor $Z_{s}$, the contribution of residue is irrelevant now. The correction of the effective mass is isotropic due to the rotation symmetry. Without loss of generalities, we assume $\mathbf{k}_s=k_{s}\mathbf{e}_{x}$ in the following. We begin with
\begin{align} \label{eq:meff2}
\partial _{k}\text{Re}\Sigma _{s}\left( \mathbf{k},0\right)
|_{\left\vert \mathbf{k}\right\vert =k_{s}}& =\text{Re}\int_{-\infty }^{\infty }%
\frac{d^2\mathbf{q}dw}{(2\pi )^{3}}  \notag \\
\times \partial _{k} \sum\limits_{r}& \frac{1}{iw+\xi _{\mathbf{k-q,}r}}V_{sr;sr}^{RPA}(\mathbf{%
k,q},iw).
\end{align}
The interaction vertex is dependent on the external momentum $\mathbf{k}$ because of the overlap of the helical eigenstates. It is instructive to consider some special cases. For weak SOC ($\gamma \ll 1$), the integration in Eq. (\ref{eq:meff2}) can be expanded to
\begin{equation}
\partial _{k}\text{Re} \Sigma_{s}\left( \mathbf{k},0\right)|_{\left\vert \mathbf{k}\right\vert=k_{s}} = \frac{k_F g}{4\pi}s\gamma +\mathcal{O}(\gamma^2).
\end{equation}
In contrast to the SOC Fermi liquid with Coulomb interaction \cite{Saraga}, our result is band dependent and has a first order correction $\gamma$. The effective mass reads as
\begin{eqnarray}\label{result:ms}
\frac{m_{s}^{\ast }}{m} = Z_s^{-1}(1+s\frac{mg}{mg+\pi}\frac{\gamma}{4\kappa})^{-1}.
\end{eqnarray}
For strong SOC ($\gamma \sim 1$), we show the numerical results along with the analytical results of Eq. (\ref{result:ms}) in Fig. \ref{figure:meff}. The effect of SOC is shown in Fig. \ref{figure:meff2}, from which we can see that the effective mass for the $s=+1$ branch is enhanced while the $s=-1$ branch is reduced with increasing strength of SOC.

\subsection{Spectral function, density of state and specific heat}

\begin{figure}[t]
\centering
\includegraphics[width=3.1in]{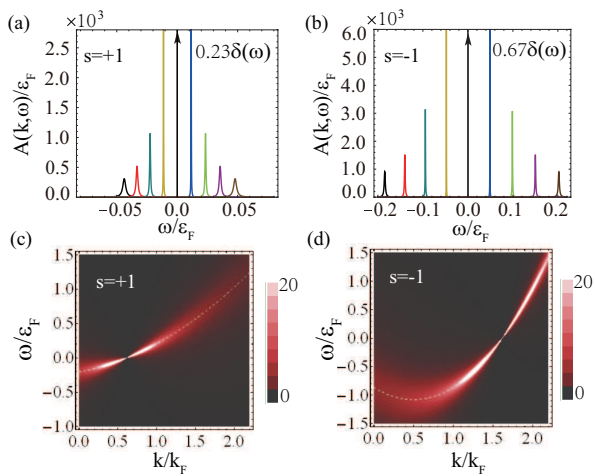}\hspace{0.5cm}
\caption{(Color online). Zero temperature spectral function at different values of $(k-k_s)/k_F$ are shown at (a) and (b). All the parameters are the same with Fig. \ref{figure:lifetime}. (a) is for $s=+1$ and (b) is for $s=-1$ respectively. The eight peaks, from left to right, correspond to $(k-k_s)/k_F=-0.01,-0.075,-0.05,-0.025,0.025,0.05,0.075,0.01$. The vertical arrows in (a) and (b) denote $\delta$ functions at $\omega=0$ with weights $0.23$ and $0.67$ respectively. (c) and (d) are density plots of the spectral function for the same parameters as above, which correspond to $s=+1$ and $s=-1$ respectively. The white dashed line denotes the modified single particle dispersion.}
\label{figure:spectrum}
\end{figure}

A close related quantity is the spectral function $A(\mathbf{k},\omega )$,
which is the imaginary part of the single particle Green's function \cite{book:Mahan,book:Fetter}
\begin{equation}
A(\mathbf{k},\omega )=-\frac{1}{\pi }\text{Im}G^{ret}(\mathbf{k},\omega ),
\end{equation}%
where $G^{ret}(\mathbf{k},\omega )$ is the retarded Green's function. It can
be straightforwardly evaluated from the time-ordered Green's function $G(%
\mathbf{k},\omega )$ as
\begin{eqnarray}\text{Im}G^{ret}(\mathbf{k},\omega )& =&\text{Im}G(\mathbf{k},\omega )%
\text{sign}(\omega ), \notag \\
\text{Re}G^{ret}(\mathbf{k},\omega )& =&\text{Re} G(\mathbf{k},\omega ).
\end{eqnarray}
With the great improvements in the spectroscopic technique, it has been possible to directly measure the low-energy spectral weight function of a 2D system in the momentum-resolved radio frequency (rf) experiments \cite{Chin2,Schunck,Veillette}.  In Fig. \ref{figure:spectrum}, we show the spectral functions of the two Rashba bands separately. The spectral functions at the Fermi surfaces as shown in Fig. \ref{figure:spectrum} (a) and (b) (vertical arrow) have the form
\begin{equation}
A(\mathbf{k}_{s},\omega )=Z_{s}\delta (\omega ).
\end{equation}
Fig. \ref{figure:spectrum} (c) and (d) are density plots of the spectral functions, which could be compared with the results of the momentum-resolved rf spectroscopy in current experiments.

Given the spectral functions, we can obtain the density of states (DOS) of the fermionic system with SOC via \cite{book:Mahan,book:Fetter}
\begin{equation} \label{eq:DOS0}
\rho(\omega)=\sum_s \rho _{s}(\omega )=\sum_s \int \frac{d^{2}\mathbf{k}}{{(2\pi )^{2}}}A_{s}(\mathbf{k}
,\omega ).
\end{equation}
For the non-interacting case, the DOS now becomes
\begin{equation}\label{eq:DOS}
\rho^0 {(\omega )}=
\begin{cases}
0, & \omega <-\frac{\kappa ^{2}k_F^2}{2m}, \\
\frac{m}{\pi}\frac{\gamma }{\sqrt{2m\omega/k_F^2+\kappa ^{2}}}, & -\frac{\kappa ^{2}k_F^2}{2m%
}<\omega <-\frac{k_F^2}{2m}, \\
\frac{m}{\pi}, & \omega >-\frac{k_F^2}{2m}.
\end{cases}
\end{equation}
The DOS is modified in the presence of $s$-wave interaction. On the Fermi surfaces, for non-interacting SOC Fermi gas, the DOS in units of $m/\pi$ is given by: $\rho^0_{+1}(E_F)=0.28$, $\rho^0_{-1}(E_F)=0.72$, and $\rho^0(E_F)=1$ for $\gamma=0.5$. For the interacting SOC Fermi gas, the DOS is evaluated numerically as follow: $\rho_{+1}(E_F)=1.37$, $\rho_{-1}(E_F)=0.83$ and $\rho(E_F)=2.20$, where all the parameters are the same with Fig. \ref{figure:lifetime}.

The quasi-particles from two helical bands both contribute to the specific heat. At low temperature, the specific heat is proportional to the DOS on the Fermi surfaces and the temperature $T$. Thus the ratio between the specific heats at low temperature is
\begin{equation} \label{eq:cvratio}
\frac{c_v}{c^0_{v}} =\frac{\rho(E_F)}{\rho^0(E_F)}=2.20,
\end{equation}
where $c^0_{v}$ is the specific heat of SOC Fermi gases without interactions.

\section{Discussions and summaries}
We have obtained various normal state quantities of 2D SOC Fermi gases in the presence of repulsive $s$-wave interaction. Ultimately, to make contact with experiments, two practical considerations warrant mention. (i): The repulsive $s$-wave interaction can be achieved on the upper branch of a Feshbach resonance. One problem should be considered is that the upper branch of a Feshbach resonance is an excited branch, and will decay to the BEC molecule state due to inelastic three-body collisions \cite{Petrov}. However, with small scattering length, the decay rate is well suppressed \cite{GBJo,DIncao} and the system may be metastable for observation. (ii): Recently, the SOC degenerate Fermi gases have been realized in the ultra-cold atom systems \cite{JZhang,Zwierlein}. The effective SOC generated in their experimental schemes is an equal weight combination of Rashba-type and Dresselhaus-type SOC. In this paper, we have investigated the case of Rashba-type SOC. The Dresselhaus-type SOC is presented in Appendix, which is demonstrated to give the same normal state properties as the Rashba case.

In current experiments, we consider the following typical experimental parameters for quasi-2D systems. One can trap about $10^4$ $^{40}$K atoms within a pancake-shaped harmonic potential with the trap frequency chosen as $2\pi \times(10,10,400)$Hz along the $(\hat{x},\hat{y},\hat{z})$ direction. The system size can be estimated as $(37.8,37.8,5.98)\mu$m. The other related parameters are taken as: $a_s=32a_0$, $\gamma=0.5$, $k_R=2\pi/\lambda=8.128 \times 10^6$m. The Fermi liquid parameters are listed in Table. \ref{tab:1} under this typical experimental setup. For comparison, we give the results of ordinary Fermi liquid and the 2DEG in semiconductors together. We should notice that the single particle spectral function captures the valuable information for low energy excitations of the Fermi liquid and can be measured by means of mometnum-resolved rf experiments \cite{HHui2,Chin2,Schunck,Veillette}. The rf spectroscopy is a technique used to probe atomic correlation by exciting atoms from occupied hyperfine states to another (usually empty) reference hyperfine state. The single-particle spectral function is obtained in experiment through the mometnum-resolved rf-transfer strength. As a result, the inferences drawn from the spectral function in the vicinity of the Fermi surfaces can be used to determine the Fermi liquid parameters describing the low energy behaviors of the normal state. Up to now, the rf experiment is the most promising method in ultracold atomic gases to measure these Fermi liquid parameters.

\begin{table}[htbp]
\caption{\label{tab:1} Normal state properties for SOC Fermi liquid ($\gamma=0.5$), ordinary Fermi liquid ($\gamma=0$), and 2DEG ($\gamma=0.051$) in semiconductor. All other parameters used here are the same with Fig. \ref{figure:lifetime}.}
 \begin{tabular}{lclcl}
  %\toprule
  \hline
  \hline
                                   & $\gamma=0.5$ & \hspace{0.2cm} $\gamma=0$ & 2DEG($\gamma=0.051$)$^{\dagger  }$ \\
  %\midrule
  \hline
\vspace{0.2cm}
   \hspace{0.3cm}$1/\tau_s ^{\dagger \dagger}$ & $0.73$kHz & $0.67$kHz & $55.36$GHz\\
\vspace{0.2cm}
   \ \ \ \ $Z_s$       & $\begin{array}{c}
                    Z_{+1}=0.23,\\
                    Z_{-1}=0.67.  \\
                 \end{array}$ & \ \ \ \  0.96 & $0.97$ \\
   \ \ $m^*_s/m$    & $\begin{array}{c}
                    m^*_{+1}/m=4.88,\\
                    m^*_{-1}/m=1.16.   \\
                 \end{array}$ & \ \ \ \ 1.04 & $0.98$ \\
  %\bottomrule
  \hline
  \hline
 \end{tabular}
\begin{itemize}
\item[$\dagger$] \small
The results for 2DEG in InGaAs are taken from Ref. \cite{Saraga}. Compared with the SOC Fermi liquid in the presence of $s$-wave interaction, the results for the 2DEG with Coulomb interaction are independent on the energy band, and the leading order correction relative to the SOC strength is $\gamma^2$.
\item[$\dagger\dagger$] \small
The values of the inverse lifetime are evaluated at $(k-k_{\pm 1})/k_F=0.01$. Compared with 2DEG in the typical semiconductor, the quasi-particle in ultracold atomic gases is much stabler.
\end{itemize}
\label{tab:1}
\end{table}
In summary, we studied the normal state properties of the SOC Fermi gas with repulsive $s$-wave interaction. The quasi-particle lifetime $\tau _{s}$, the renormalization factor $Z_{s}$, the effective mass $m^{\ast }_s/m$ are calculated, which embody the main properties of a quasi-particle. To make contact with experiments directly, we calculated the spectral function $A(\mathbf{k},\omega)$, density of state $\rho(\omega)$, and the specific heat $c_{v}$ at low temperature. These quantities provide a good description of the low energy physics with SOC and $s$-wave interaction, which are measurable in current experiments. The analytical and numerical results show that the normal state properties are distinct for the two energy bands, and the leading correction relative to the ordinary Fermi liquid with $s$-wave interaction is on the order of $\gamma$, which are strikingly different from the SOC Fermi liquid with Coulomb interaction \cite{Saraga}. We expect our microscopic calculations of the Fermi liquid parameters and related quantities would have the immediate applicability to the SOC Fermi gases in the upper branch of the energy spectrum.

\begin{acknowledgments}
We acknowledge helpful discussions with Jinwu Ye, Han Pu, Congjun Wu, and Hui Hu. This work was supported by the NKBRSFC under Grants No. 2009CB930701, No. 2010CB922904, No. 2011CB921502, and No. 2012CB821300, NSFC under Grants No. 10934010, and NSFC-RGC under Grants No. 11061160490 and No. 1386-N-HKU748/10.
\end{acknowledgments}

\appendix
\section{Relationships between the Rashba SOC and Dresselhaus SOC}
In this paper, the normal state properties of the Fermi liquid with Rashba-type SOC
is studied. Now, we would like to demonstrate that another type of SOC, namely the Dresselhaus-type,
gives exactly the same results for the normal state properties considered
here.

We start with the single particle Hamiltonian with Dresselhaus SOC \cite{NIST4}
\begin{equation}
\mathcal{H}_{D}=\frac{\mathbf{p}^{2}}{2m}+
\alpha(-p_{y}\sigma _{x}-p_{x}\sigma _{y})-\mu .
\end{equation}
The helicity bases with Dresselhaus SOC are
\begin{equation}
|\mathbf{p},s\rangle _{D}=-\frac{1}{\sqrt{2}}\left(
\begin{array}{c}
1 \\
ise^{-i\phi \left( \mathbf{p}\right) }
\end{array}%
\right) ,s=\pm 1,
\end{equation}
where the subscript $D$ represents the Dresselhaus-type SOC. The starting point of the microscopic calculation is the self-energy $\Sigma _{s}$ for each band. To illustrate the relationships of the two types of SOC, it is essential to derive the relationships of the Feynman rules between the two cases. For the Dresselhaus-type SOC, the non-interacting Green's function and the interaction vertex in the helicity bases reads
\begin{eqnarray}
G_{s}^{D}\left(\mathbf{k}, \omega\right)  &=&\frac{1}{\omega
-\xi _{\mathbf{k,}s}+i \text{sgn}(\omega)0^+}, \\
V_{ss^{\prime };rr^{\prime }}^{D}(\mathbf{k,p,q}) &=&gf_{ss^{^{\prime
}}}^{D}(\theta _{\mathbf{k}},\theta _{\mathbf{k+q}})f_{rr^{^{\prime
}}}^{D}(\theta _{\mathbf{p}},\theta _{\mathbf{p-q}}).
\end{eqnarray}
The energy
spectrum of the Dresselhaus-type SOC is the same with the Rashba-type. Thus the single particle retarded Green's
function within the Dresselhaus representation is the same with the result for Rashba-type SOC given in Eq. (\ref{eq:G0}). The overlap factor $f^D_{ss^{^{\prime }}}(\theta _{
\mathbf{k}},\theta _{\mathbf{k+q}})$ for the Dresselhaus-type SOC is given
by
\begin{eqnarray}
f_{ss^{^{\prime }}}^{D}(\theta _{\mathbf{k}},\theta _{\mathbf{k+q}})
=\frac{1}{2}(1+ss^{^{\prime }}e^{-i[\phi \left( \mathbf{k}\right) -\phi
\left( \mathbf{k+q}\right) ]}).
\end{eqnarray}
Compared with Eq. (\ref{eq:overlap}), we find the overlap factors are conjugated for the two cases \begin{eqnarray}
f_{ss^{^{\prime }}}^{D}(\theta _{\mathbf{k}},\theta _{\mathbf{k+q}
})=f_{ss^{^{\prime }}}^{R}(\theta _{\mathbf{k}},\theta _{\mathbf{k+q}
})^{\ast }.
\end{eqnarray}

The bare susceptibility in Matsubara formalism is given by
\begin{equation}
\chi^D (\mathbf{q,}i\omega _{n})\!=\!k_{B}T\!\!\sum_{\substack{\mathbf{
k,}i\omega _{m} \\ s,r}}\!\!G^D_{s}(\mathbf{k,}i\omega _{m})G^D_{r}(\mathbf{k-q,}i\omega
_{m}-i\omega _{n})\mathcal{F}^D_{sr},
\end{equation}
where the factor
\begin{equation}
\mathcal{F}^D_{sr}=\frac{1+sr\cos \theta }{2}
\end{equation}
is the same with the Rashba case in Eq. (\ref{eq:factor}). So the bare susceptibility $\chi^D$ for the Dresselhaus type SOC is equal to the case for the Rashba type SOC. The self-energy is given by Eq. (\ref{residue}) and (\ref{imagline}). The integrand function includes the following factors: the single particle's Green's function, the bare susceptibility $\chi^D(\mathbf{q,}i\omega _{n})$ and the overlap factor $\mathcal{F}^D_{sr}$, which are all demonstrated to be the same for the two types of SOC. Therefore we conclude that the normal state properties calculated here, such as the quasi-particle lifetime $\tau _{s}(\mathbf{k})$, the renormalization factor $Z_{s}$ and the effective mass $m^{\ast }_s/m$ are all the same exactly for the two types of SOC.

\section{spectral function in the spin representation}
In this work, we studied the microscopic parameters such as $\tau_{\mathbf{k},s}$, $Z_s$ and $m^*_s/m$ and associated experimental observable such as $A_s(\mathbf{k},\omega)$, $\rho(E_F)$ and $c_v$ in the helicity representation. Theoretically speaking, the helicity bases provides a clear representation for the description of the microscopic picture of the single particle excitation of the repulsive SOC Fermi gas. In the normal state regime, the quasi-particle in the weakly repulsive SOC Fermi gas can be adiabatically connected with the particle with helicity in the non-interacting theory. In this Appendix, we will also present the main results in the spin representation in our manuscript, since experimentalists find it handier to work with spin representation. In the following, we will transform the results to the spin representation.

In the spin representation, the Green's functions of the SOC Fermi gas has the of $2 \times 2$ matrix form. The non-interacting form is given by
\begin{equation}
G_0(\mathbf{k},\omega)_{\alpha,\beta}= \sum_s \frac{1}{\omega-\xi_{\mathbf{k},s}+i \text{sgn}(\omega) 0^+}P_s(\mathbf{k}) ,
\end{equation}
where $P_s (\mathbf{k})=[1+s (\hat{\mathbf{z}}\times \hat{\mathbf{k}}) \cdot \mbox{\boldmath$\sigma$}]/2$ is the projection operator to the helicity bases. Considering the $s$-wave interaction, there is a modification of the quasi-particle dispersion $\xi^*_{\mathbf{k},s}$, and a finite lifetime of the quasi-particle $\tau_{\mathbf{k},s}$ corresponding to the imaginary part of the self-energy. In the weakly repulsive regime, the polarization of the quasi-particle is hold because of the stability from its non-trivial topology structure, such that we obtain the many-body Green's function as
\begin{equation}
G_{\alpha,\beta}(\mathbf{k},\omega)= \sum_s \frac{1}{\omega-\xi^*_{\mathbf{k},s}-i \text{sgn}(\omega) \frac{\Gamma_{\mathbf{k},s}}{2}}P_s(\mathbf{k}).
\end{equation}
Based on this form of Green's function modified with $s$-wave interaction, we could obtain the spectral function in the spin representation,
\begin{equation}
A_{\alpha,\beta}(\mathbf{k},\omega)= \sum_s P_s (\mathbf{k})_{\alpha,\beta} A_s(\mathbf{k},\omega),
\end{equation}
where $A_s$ is the spectral function in the helicity bases representation. We could find that
\begin{eqnarray}
A_{\uparrow,\uparrow}(\mathbf{k},\omega)&=& A_{\downarrow,\downarrow}(\mathbf{k},\omega)= \frac{1}{2}(A_+(\mathbf{k},\omega)+A_-(\mathbf{k},\omega)), \\
A_{\uparrow,\downarrow}(\mathbf{k},\omega)&=&A_{\downarrow,\uparrow}(\mathbf{k},\omega)^* \notag \\
&=&-\frac{i e^{-i \theta_{\mathbf{k}}}}{2}(A_+(\mathbf{k},\omega)-A_-(\mathbf{k},\omega)).
\end{eqnarray}
The elements $A_{\uparrow,\uparrow}$, $A_{\downarrow,\downarrow}$ are real and equal to each other, while the off-diagonal elements $A_{\uparrow,\downarrow}$, $A_{\uparrow,\downarrow}$ have an angle dependence $e^{-i \theta_{\mathbf{k}}}$, $e^{i \theta_{\mathbf{k}}}$, which is the very effect of spin-orbit coupling.

The other qualities such as the density of states at the Fermi energy, and the low temperature specific heat are both independent to the representation we use. Therefore they could be compared with experiment straightforwardly.

\end{document}